\documentclass[11pt]{article}
\usepackage{xspace}
\usepackage{graphicx}
\usepackage{amsmath}
\usepackage{amssymb}
\usepackage{color}

\definecolor{Red}{rgb}{1,0,0}
\definecolor{Green}{rgb}{0,1,0}
\definecolor{Blue}{rgb}{0,0,1}
\definecolor{Black}{rgb}{0,0,0}

\def\beq{\begin{equation}}
\def\eeq#1{\label{#1}\end{equation}}
\def\eeqn{\end{equation}}

\def\beqa{\begin{eqnarray}}
\def\eeqa#1{\label{#1}\end{eqnarray}}
\def\eeqan{\end{eqnarray}}

\let\bar=\overbar

\def\Dslash{\not{\hbox{\kern-4pt $D$}}}
\def\dslash{\not{\hbox{\kern-2pt $\del$}}}

\def\msb{{\bar{\ssstyle M \kern -1pt S}}}

\def\Title#1{\begin{center} {\Large {\bf #1} } \end{center}}

\begin{document}

\Title{Latest Results from the Alpha Magnetic Spectrometer on the International Space Station}

\bigskip\bigskip


\begin{raggedright}  

Martin Pohl\index{Pohl, M.}, {\it DPNC and CAP Gen\`eve, University of Geneva}\\

\begin{center}\emph{On the behalf of the AMS Collaboration.}\end{center}
\bigskip
\end{raggedright}

{\small
\begin{flushleft}
\emph{To appear in the proceedings of the Interplay between Particle and Astroparticle Physics workshop, 18 -- 22 August, 2014, held at Queen Mary University of London, UK.}
\end{flushleft}
}

\begin{abstract}
I review latest results from AMS on electrons and positrons in primary cosmic rays in the GeV to TeV energy range. Separate fluxes for electrons and positrons as well as their sum are presented. Neither of the fluxes is compatible with a simple power law. New data on the fraction of positrons in the joint electron and positron flux are also presented, which extend the energy range of our previous observation and increase its precision. The new results show, for the first time, that above about 200 GeV the positron fraction no longer exhibits an increase with energy.  The results confirm that a common new source of electrons and positrons exists.
\end{abstract}

\section{Introduction}

The mission of the Alpha Magnetic Spectrometer (AMS) on board of the International Space Station (ISS) is to establish a complete inventory of cosmic rays in Near Earth Orbit in the GeV to TeV energy range. Since all cosmic ray spectra fall by roughly three orders of magnitude every decade in energy, the product of acceptance and exposure time not only determines the statistical accuracy of the result, but also the energy reach. Thus size matters, but also redundancy in measuring crucial parameters is of prime importance~\cite{TIPP2014}. AMS measures the spectrum as well as the chemical and isotopic composition of cosmic rays including rare components like electrons, positrons, antiprotons and heavy nuclei. The aim is to understand astrophysical sources of cosmic rays, their acceleration and transport mechanisms inside our galaxy. Rare components like positrons may also reveal non-astronomical sources of cosmic rays, like the self-annihilation or decay of dark matter. Consequently, there has been a strong interest in the cosmic ray positron fraction in both particle physics and astrophysics~\cite{DM_theory}. The positron fraction is defined as the ratio of the positron flux to the combined flux of positrons and electrons. The first results from AMS on the positron fraction were reported in~\cite{AMS_pf_250}. They generated widespread interest~\cite{DM_fits}. 

Here I review new results~\cite{AMS_pf_500, AMS_e7p5,AMS_alle1} based on all the data collected during 30 months of AMS operations on the ISS from 19 May 2011 to 26 November 2013. Due to the excellent and steady performance of the detector, an increase of the data sample by a factor of 1.7 is obtained with respect to the first publication~\cite{AMS_pf_250}. The electron flux in the energy range from 0.5 to 700 GeV as well as the positron flux from 0.5 to 500 GeV are reported separately. The joint electron and positron flux is measured up to 1 TeV and benefits from reduced systematics. For the positron fraction, the energy range is extended up to 500 GeV and the precision increased. 

\section{The Alpha Magnetic Spectrometer in the ISS}
\begin{figure}[!ht]
\begin{center}
\includegraphics[width=0.4\columnwidth]{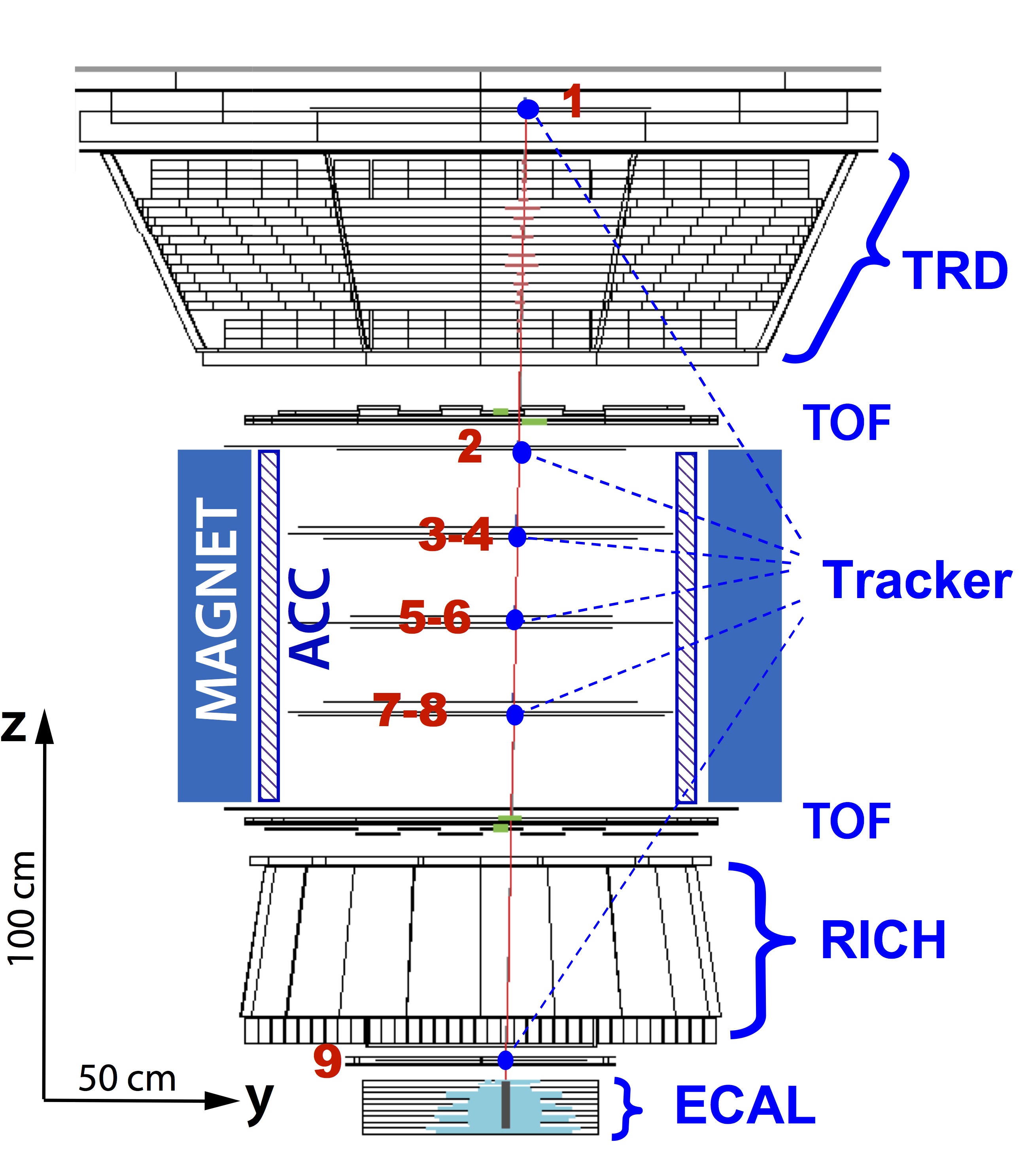}
\caption{A 369 GeV positron event as measured by the AMS detector on the ISS in the bending (y-z ) plane. Tracker planes 1 to 9 measure the particle charge, sign, and momentum. The TRD identifies the particle as e$^\pm$. The TOF measures the absolute charge value to be 1 and ensures that the particle is downward-going. The RICH independently measures the charge and velocity. The ECAL measures the 3D shower profile, independently identifies the particle as an e± and measures its energy. A positron is identified by 1) positive rigidity in the tracker, 2) an e$^\pm$ signal in the TRD, 3) an e$^\pm$ signal in the ECAL and 4) the matching of the ECAL shower energy and axis with the momentum measured with the tracker and magnet.}
\label{fig:positron_event}
\end{center}
\end{figure}

The layout of the AMS-02 detector~\cite{AMS_detector} is shown in Fig.~\ref{fig:positron_event}. It consists of 9 planes of precision silicon tracker with two outer planes, 1 and 9, and the inner tracker, planes 2-8; a transition radiation detector, TRD; four planes of time of flight counters, TOF; a permanent magnet with a central field strength of 0.15 T; an array of anti-coincidence counters, ACC, inside the magnet bore; a ring imaging Cherenkov detector, RICH; and an electromagnetic calorimeter, ECAL. The figure also shows a high energy positron of 369 GeV recorded by AMS. AMS operates without interruption on the ISS and is monitored continuously from the ground. The maximum detectable rigidity $p/Z$ over tracker planes 1 through 9, with a lever arm of 3 m, is about 2 TV. Detector performance is steady over time.

Three main detectors provide clean and redundant identification of positrons and electrons with independent suppression of the proton background. These are the TRD (above the magnet), the ECAL (below the magnet) and the tracker. The TRD and the ECAL are separated by the magnet and the tracker. This ensures that most of the secondary particles produced in the TRD and in the upper TOF planes are swept away and do not enter into the ECAL. Events with large angle scattering are also rejected by a quality cut on the measurement of the trajectory using the tracker. The matching of the ECAL energy, $E$, and the momentum measured with the tracker, $p$, greatly improves the proton rejection. To differentiate between e$^\pm$ and protons in the TRD, signals from the 20 layers of proportional tubes are combined in a TRD estimator formed from the ratio of the log-likelihood probability of the e$^\pm$ hypothesis to that of the proton hypothesis in each layer. The proton rejection power of the TRD estimator at 90\% e$^\pm$ efficiency measured on orbit is $10^3$ to $10^4$~\cite{AMS_pf_250}. To cleanly identify electrons and positrons in the ECAL, an estimator, based on a Boosted
Decision Tree algorithm~\cite{BDT} is constructed using the 3D shower shape in the ECAL. The proton rejection power of the ECAL estimator reaches $10^4$ when combined with the energy/momentum matching requirement $E/p > 0.75$~\cite{AMS_pf_250}. The entire detector has been extensively calibrated in a test beam at CERN with e$^+$ and e$^-$ from 10 to 290 GeV, with protons at 180 and 400 GeV, and with $\pi^\pm$ from 10 to 180 GeV, which produce transition radiation equivalent to protons up to 1.2 TeV. A Monte Carlo program based on the Geant 4.9.4 package~\cite{GEANT4} is used to simulate physics processes and signals in the detector.

\section{Electron and Positron Fluxes}
The omnidirectional fluxes of cosmic ray electrons and positrons in the energy bin $E$ of width $\Delta E$ are given by:
\begin{equation}\label{eq:flux}
\Phi_{\mathrm{e}^\pm}(E) = \frac{N_{\mathrm{e}^\pm}(E)}{A_{\mathrm{eff}} \cdot \epsilon_{\mathrm{trig}} \cdot T(E) \cdot \Delta E} 
\end{equation}
where $N_{\mathrm{e}^-}$ is the number of electrons, $N_{\mathrm{e}^+}$ is the number of positrons, $A_{\mathrm{eff}}$ is the effective
acceptance,  $\epsilon_{\mathrm{trig}}$ is the trigger efficiency, and $T$ is the exposure time. The effective acceptance is defined as the product between the geometric acceptance of about 550 cm$^2$ sr, the selection efficiency for well measured events and the identification efficiency for electrons and positrons. This product is determined by Monte Carlo simulation and receives a minor correction taking into account differences in the efficiencies between data and Monte Carlo simulation. The trigger efficiency is determined from data using unbiased triggers, it is found to be 100\% above 3 GeV. The selection efficiency is determined from the Monte Carlo simulation
and found to be a smooth function of energy with a value of  about 70\% at 100 GeV. The exposure time $T(E)$ is counted taking into account the lifetime at each location as well as the local geomagnetic cut-off, excluding time spent in the South Atlantic Anomaly; it amounts to $1.4 \times 10^7$s at 5 GeV, $3.4 \times 10^7$s at 10 GeV and grows to a constant $6.1 \times 10^7$s above 30 GeV. 

Signals from the 17 radiation length ECAL are scaled to provide the incoming (top of AMS) energy, $E$, of electrons and positrons. In the beam tests of the AMS detector, the energy resolution has been measured to be $\sigma(E)/E =\sqrt{(0.104)^2/E + (0.014)^2}$ with $E$ in GeV. The absolute energy scale is verified by using minimum ionizing particles and the ratio $E/p$. These results are compared with the test beam values where the beam energy is known to high precision. This comparison limits the uncertainty of the absolute energy scale to 2\% in the range covered by the beam test results, 10 to 290 GeV. Below 10 GeV it increases to 5\% at 0.5 GeV and above 290 GeV to 4\% at 700 GeV. This is treated as an uncertainty of the bin boundaries. The bin widths, $\Delta E$, are chosen to be at least two times the energy resolution to minimize migration effects. The bin-to-bin migration error is about 1\% at 1GeV; it decreases to 0.2\% above 10 GeV. With increasing energy the bin width is smoothly widened to ensure adequate statistics in each bin.

The identification of the e$^-$ and e$^+$ signal requires rejection of the proton background. Cuts are applied on the $E/p$ matching and the reconstructed depth of the shower maximum. This makes the negatively charged sample, as determined by the rigidity, a sample of pure electrons. A cut on the ECAL estimator is applied to further reduce the proton background in the positive rigidity sample after which the numbers of positrons and protons are comparable at all energies. The identification efficiency, $\epsilon_{\mathrm{id}}$ is defined using the Monte Carlo simulation as the efficiency for electrons to pass these three cuts. It is identical for both
electrons and positrons. In order to correct for small differences between data and simulation, a negative rigidity sample is selected for every cut using information from the detectors unrelated to that cut. The effects of the cut are compared between data and Monte Carlo simulation. The resulting correction is found to be a smooth, slowly falling function of energy. It is -2\% at 10 GeV and -6\% at 700 GeV relative to $\epsilon_{\mathrm{id}}$. In each energy bin, a template fit to the discriminant variables determines the number of electrons, positrons and protons in the sample. After a correction for charge confusion obtained from Monte Carlo and checked with data (see below), one obtains $N_{\mathrm{e}^-}$ and $N_{\mathrm{e}^+}$. 

In total, 9.23 million events are identified as electrons and 0.58 million as positrons. These numbers are slightly less than the numbers used below to determine the positron fraction due to tighter selection criteria (such as on the exposure time) used to minimize the uncertainty of the separate flux measurements.

The systematic error associated with the uncertainty of the template shapes for the signal and the background is due to the finite accuracy of the TRD alignment and calibration as well as to the statistics of the data samples used to construct the templates. This is the leading contribution to the total systematic error above 300 GeV. The amount of charge confusion is well reproduced by the Monte Carlo simulation and a systematic uncertainty takes into account the small differences between data and the Monte Carlo simulation. This uncertainty is only significant for $N_{\mathrm{e}^+}$ in the highest energy bin. The systematic error on the effective acceptance is given by the uncertainties on the correction for differences between data and simulation, derived from their comparison for ever cut. This error includes an overall scaling uncertainty of 2\% which introduces a correlation between energy bins and between the electron and positron fluxes. The acceptance uncertainty is the leading contribution to the systematic error below 300 GeV. The total systematic error is taken as the quadratic sum of these three contributions and
the minute bin-to-bin migration systematic. As an example, in the energy bin from 59.1 to 63.0 GeV, the statistical error on the positron flux is 4.9\% and the total systematic error is 2.9\% with 0.8\% from the TRD templates, 0.4\% from charge confusion, 2.8\% from the
effective acceptance, and 0.2\% from bin-to-bin migration. Large variations of the cuts have been applied to verify the above error assessment. The time stability of the result has also been verified. 

\begin{figure}[!ht]
\begin{center}
\includegraphics[width=0.48\columnwidth]{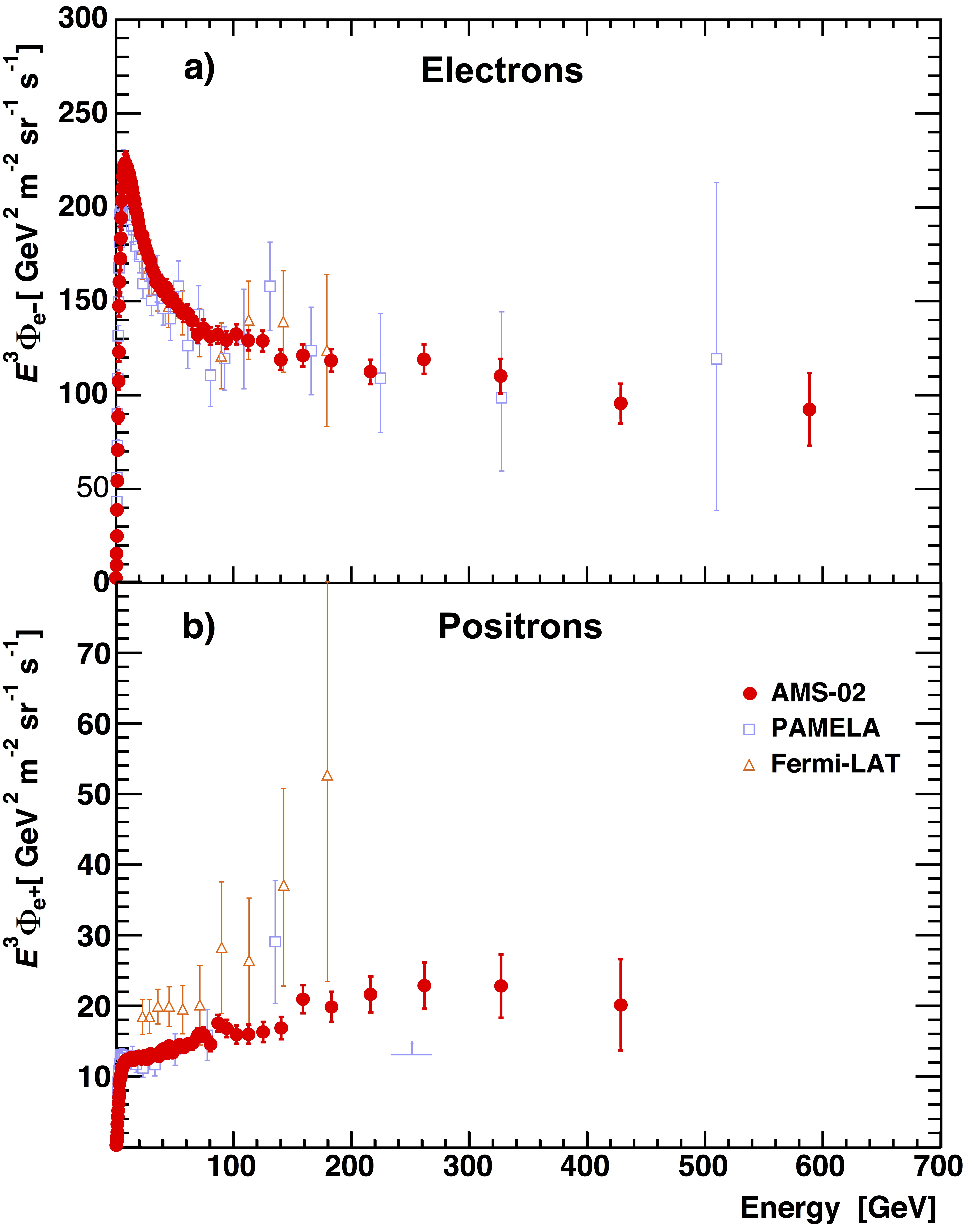} \hfill
\includegraphics[width=0.48\columnwidth]{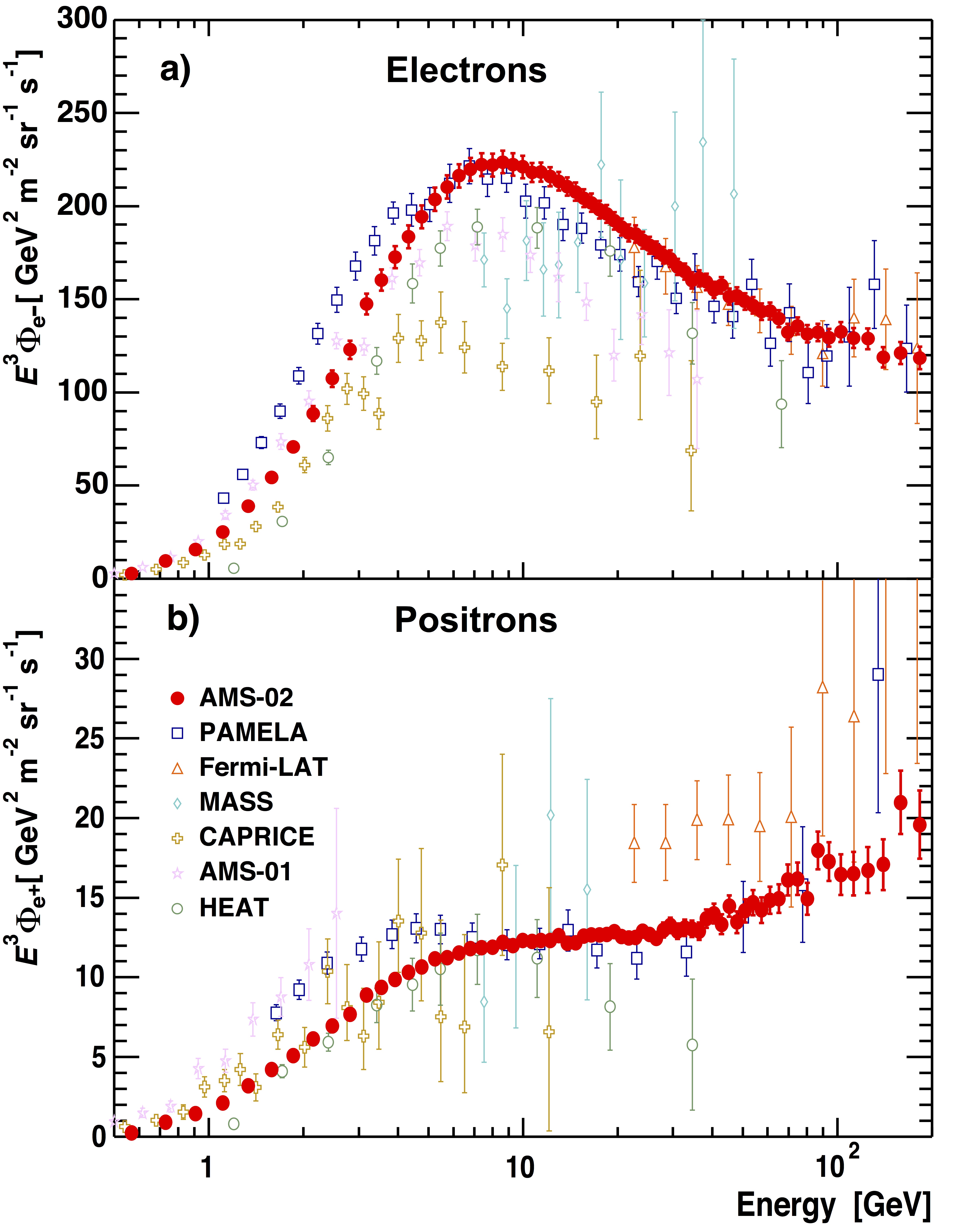}
\caption{The AMS (a) electron and (b) positron fluxes, multiplied by $E^3$ (red points). Statistical and systematic
uncertainties of the AMS results have been added in quadrature. Also shown are the most recent
measurements from PAMELA~\cite{Pamela} and Fermi-LAT~\cite{Fermi-LAT}. The right hand plots zoom in on the region below 200 GeV for better comparison with previous results~\cite{Others_separate}.}
\label{fig:ep_flux}
\end{center}
\end{figure}

The electron and positron fluxes multiplied by $E^3$ are presented in Fig.~\ref{fig:ep_flux} together with the most recent measurements~\cite{Pamela,Fermi-LAT} for comparison. The figure also shows the detailed behavior for both electrons and positrons below 200 GeV together with previous measurements~\cite{Pamela,Fermi-LAT,Others_separate}  in this energy range. Below about 10 GeV, the behavior for both electrons and positrons is affected by solar modulation as seen in our data by variations of the fluxes over the data taking
interval. However, above about 20 GeV the effects of solar modulation are insignificant within the current experimental accuracy. The data show that above 20 GeV and up to 200 GeV the electron flux decreases more rapidly with energy than the positron flux, that is, the
electron flux is softer than the positron flux. This is not consistent with only the secondary production of positrons~\cite{prop_models}.
Neither the electron flux nor the positron flux can be described by single power laws ($\propto E^{-\gamma}$) over the entire range. Power law fits over different energy ranges show that $\gamma_{\mathrm{e}^+}$ hardens from $2.97 \pm 0.03$ (fit from 15.1 to 31.8 GeV) to $2.75 \pm 0.05$ (fit from 49.3 to 198 GeV). Correspondingly, $\gamma_{\mathrm{e}^-}$ hardens from
$3.28 \pm 0.03$ (fit from 19.0 to 31.8 GeV) to $3.15 \pm 0.04$ (fit from 83.4 to 290 GeV) and then levels off.  Above about 200 GeV, $\gamma_{\mathrm{e}^+}$ exhibits a tendency to soften with energy. This is consistent with our observation (see below) that above 200 GeV the positron fraction, i.e.~the ratio of positron to electron plus positron flux, is no longer increasing with energy.
 
When forming the sum of electron and positron fluxes systematics associated to charge confusion cancel.  The flux is thus evaluated according to Equ.~\ref{eq:flux} for the sum of electrons and positrons, in 74 energy bins from 0.5 GeV to 1 TeV. The bin width is chosen to be at least twice the energy resolution, the migration error is about 1\% at 1 GeV decreasing to 0.2\% above 10 GeV. The absolute energy scale is verified as described above. Events are selected requiring the presence of a downward-going, $\beta > 0.83$ particle which has hits in at least 8 of the 20 TRD layers and a single track in the tracker passing through the ECAL. Events with an energy deposition compatible with a minimum ionizing particle in the first $5X_0$ of the ECAL are rejected. Events with $|Z| > 1$ are rejected using $dE/dx$
in the tracker and TRD. Secondary particles of atmospheric origin are rejected with cutoff requirement discussed above.
In each energy bin, TRD classifier templates of the (e$^+$+e$^-$) signal and the proton background are constructed from the data using pure samples of e$^-$ and protons. These samples are selected using the ECAL estimator, $E/p$ matching  and the charge-sign. The templates are evaluated separately in each bin, however the signal templates show no dependence on the energy above about 10GeV. Therefore, all the e$^-$ selected in the range 15.1 to 83.4 GeV are taken as a unique signal template up to the highest energies. The sum of the signal and background templates is fit to the data by varying their normalizations. This yields the number of signal (e$^+$+e$^-$) events, 10.6 million in the full energy range, the number of background (proton) events, and the statistical errors on these numbers. The other ingredients to the flux measurement are determined analogously to the separate flux measurements discussed above.  

\begin{figure}[!ht]
\begin{center}
\includegraphics[width=0.68\columnwidth]{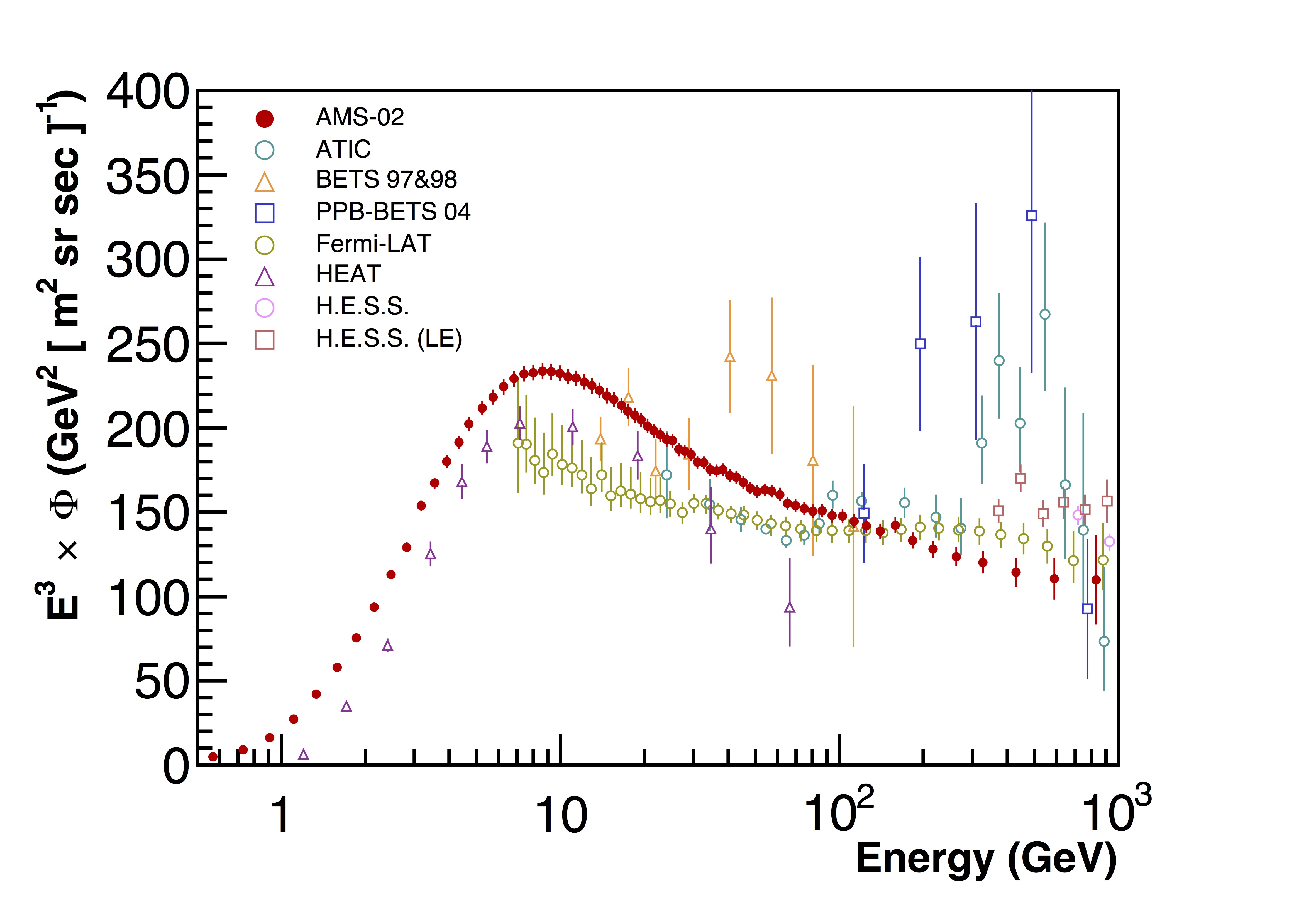} 
\caption{The flux of electrons plus positrons measured by AMS (red points) multiplied by $E^3$
versus energy. The AMS error bars are the quadratic sum of the statistical and systematic errors.
Also shown are the results from earlier experiments~\cite{Others_sum}.}
\label{fig:ep_sum}
\end{center}
\end{figure}

Fig.~\ref{fig:ep_sum} shows the result of this analysis. A major experimental advantage of the combined flux analysis compared to the measurement of the individual positron and electron fluxes, particularly at high energies, is that the selection does not depend on the charge-sign. Another advantage is that it has a higher overall efficiency. Consequently, this measurement is extended to 1TeV with less
overall uncertainty over the entire energy range. Systematic uncertainties arise from (i) the event selection, (ii) the acceptance, and (iii) bin-to-bin migration and are evaluated as described above. As seen in Fig. 3, the flux cannot be described by a single power law over the entire range. The lowest starting energy of a sliding window that gives consistent spectral indices at the 90\% C.L. for any boundary yields a lower limit of 30.2 GeV, above which we find $\gamma = 3.170 \pm 0.008 \pm 0.008$, where the first error is the combined statistical and systematic uncertainty and the second error is due to the energy scale uncertainty.  It is important to note that a single power law can describe the electron flux above 52.3 GeV; a single power law, with a different spectral index, can describe the positron flux above 27.2 GeV. 

\section{Positron Fraction}

When forming the positron fraction $\Phi_{\mathrm{e}^+}/(\Phi_{\mathrm{e}^+}+\Phi_{\mathrm{e}^-})$, the analysis simplifies since systematics corresponding to acceptance and efficiencies cancel to a large extent. With respect to our previous publication~\cite{AMS_pf_250}, systematic errors have decreased with increasing statistics in the high energy region. As other uncertainties have decreased, the contribution of the absolute energy scale uncertainty became noticeable, as quantified above. It results, however, in a negligible contribution to the total systematic error, except below 5GeV, where it dominates.

In each energy bin, the 2-dimensional templates for e$^\pm$ and the background are fit to data in the [TRD estimator-$\log(E/p)$] plane by varying the normalizations of the signal and the background. This method provides a data-driven control of the dominant
systematic uncertainties by combining the redundant and independent TRD, ECAL, and tracker information. The templates are determined from high statistics electron and proton data samples selected using tracker and ECAL information including charge sign,
track-shower axis matching, and the ECAL estimator. The purity of each template is verified using Monte Carlo simulation.
The fit is performed simultaneously for the positive and negative rigidity data samples in each energy bin yielding the number of positrons, the number of electrons, the number of protons, and the amount of charge confusion, where charge confusion is defined as the
fraction of electrons or positrons reconstructed with a wrong charge sign. 

From the bin-by-bin fits, the sample contains $10.9 \times 10^6$ primary positrons and electrons and $3.50 \times 10^6$ protons. A total of $0.64 \times 10^6$ events are identified as positrons. There are several systematic uncertainties. In addition to the energy scale, bin-to-bin migration, and a slightly asymmetric acceptance of e$^+$ and e$^-$ below 3 GeV, there are also the systematic uncertainties from event selection, charge confusion, and the templates. To evaluate the systematic uncertainty related to event selection, the complete analysis is repeated in every energy bin over 1000 times with different cut values, such that the selection efficiency varies up to 30\%. The distribution of the positron fraction resulting from these 1000 analyses contains both statistical and systematic effects. The difference between the width of this distribution from data and from Monte Carlo simulation quantifies this
systematic uncertainty.

Two sources of charge confusion dominate. The first source is related to the finite resolution of the tracker and multiple scattering. It is mitigated by the $E/p$ matching and quality cuts of the trajectory measurement including the track $\chi^2$, charge measured in the tracker and charge measured in the TOF. The second source is related to the production of secondary tracks along the path of the primary e$^\pm$ in the tracker. It was studied using control data samples of electron events where the ionization in the lower TOF counters corresponds to at least two traversing particles. Both sources of charge confusion are found to be well reproduced by the Monte Carlo simulation and the respective templates are derived from the Monte Carlo. The systematic uncertainties due to these two effects are obtained by varying the background normalizations within the statistical limits and comparing the results with the Monte Carlo simulation. They were examined in each energy bin. The proton contamination in the region populated by positrons is small. It is accurately
measured using the TRD estimator. The amount of proton contamination has a negligible contribution to the statistical error.
The systematic error associated with the uncertainty of the data derived templates arises from their finite statistics. It is measured by varying the shape of the templates within the statistical uncertainties. Its contribution to the overall error is small compared
to the statistical uncertainty of data and is included in the total systematic error.

\begin{figure}[!ht]
\begin{center}
\includegraphics[width=0.68\columnwidth]{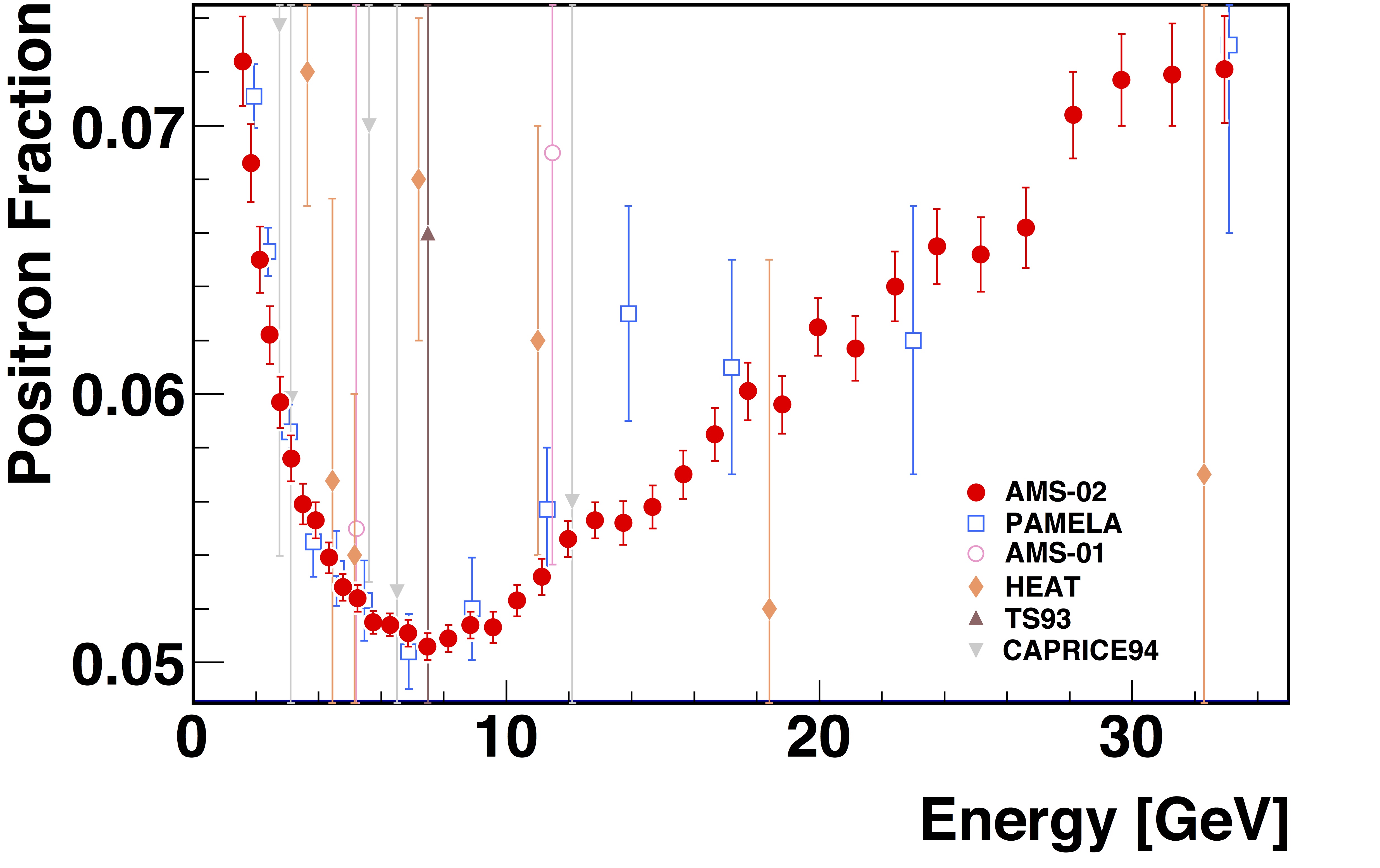} 
\caption{The positron fraction from 1 to 35 GeV. It shows a rapid decrease from 1 to about 8GeV followed by a steady increase. The AMS data (red points) provide accurate information on the minimum of the positron fraction.}
\label{fig:pfrac_35}
\end{center}
\end{figure}
  
\begin{figure}[!ht]
\begin{center}
\includegraphics[width=0.68\columnwidth]{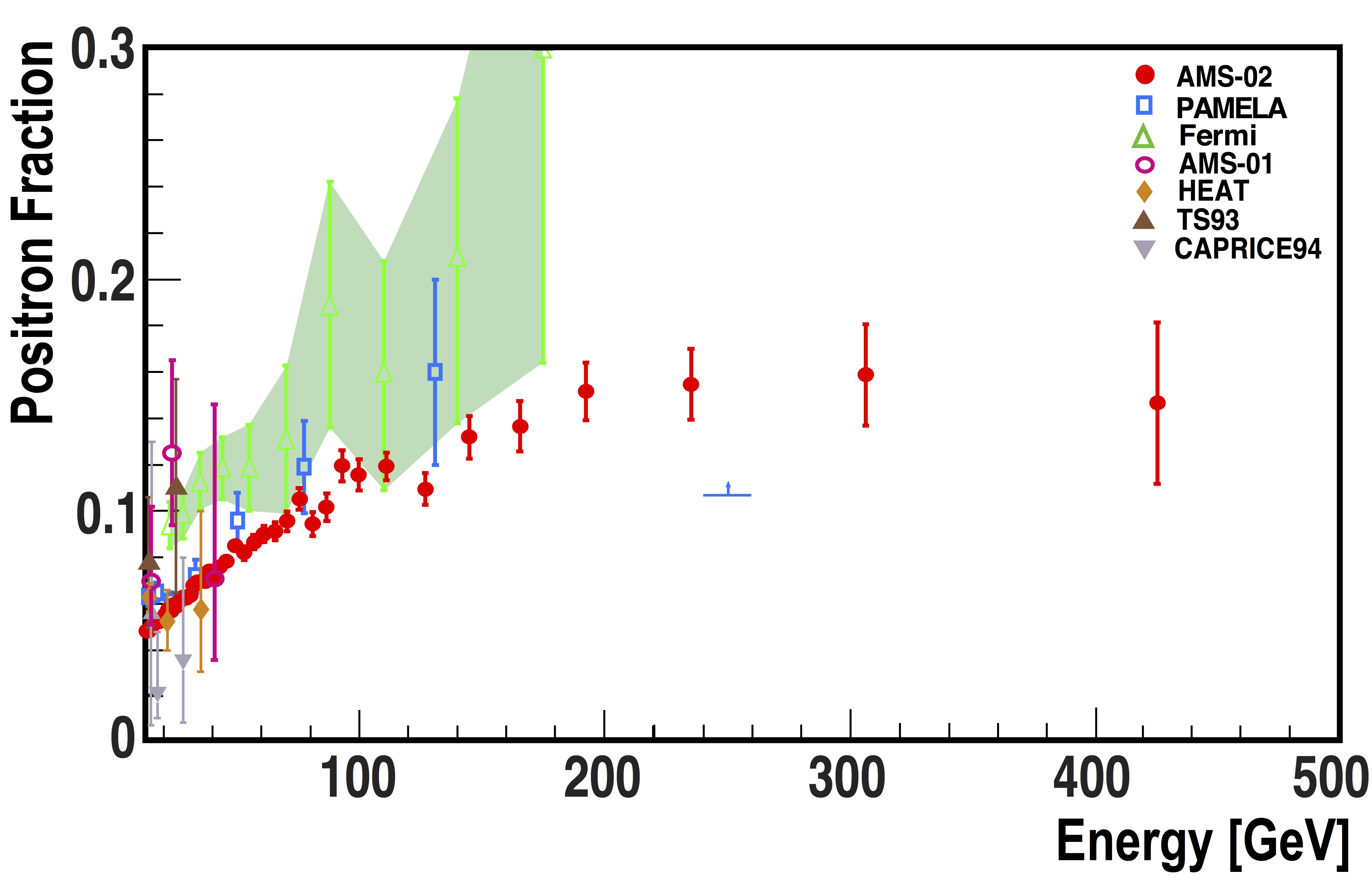} 
\caption{The positron fraction above 10 GeV, where it begins to increase. The present AMS measurement (red points) extends the energy range to 500 GeV and demonstrates that above about 200GeV the positron fraction is no longer increasing. Measurements from PAMELA~\cite{Pamela,Pamela_pfrac} (the horizontal blue line is their lower limit), Fermi-LAT~\cite{Fermi-LAT}, and other experiments~\cite{TS93,Wizard,HEAT,AMS-01} are also shown.}
\label{fig:pfrac_500}
\end{center}
\end{figure}

Fig.~\ref{fig:pfrac_35} shows the behavior of the positron fraction at low energies, from 1 to 35 GeV. As seen, below about 8 GeV the positron fraction decreases rapidly as expected from the diffuse production of positrons~\cite{prop_models}. Then the fraction begins to increase steadily with energy. The AMS data provide accurate information on the minimum of the positron fraction.
Our earlier result~\cite{AMS_pf_250} in which we observed the increase of the positron fraction with
decreasing slope above 20 GeV, is consistent with this new measurement. The increase of the
positron fraction has been reported by earlier experiments: TS93~\cite{TS93},Wizard/CAPRICE~\cite{Wizard},
HEAT~\cite{HEAT}, AMS-01~\cite{AMS-01}, PAMELA~\cite{Pamela,Pamela_pfrac}, and Fermi-LAT~\cite{Fermi-LAT}.
The new result extends the energy range to 500 GeV and is based on a significant increase
in the statistics by a factor of 1.7. Fig.~\ref{fig:pfrac_500} explores the behavior of the positron fraction at high energies ($>10$ GeV) and compares it with earlier measurements. We observe that above about 200 GeV the positron fraction is no longer increasing with energy.

To examine the energy dependence of the positron fraction quantitatively in a model independent way, straight line fits were performed over the entire energy range with a sliding energy window, where the width of the window varies with energy to have sufficient sensitivity
to the slope. Each window covers about 8 bins, at energies above 200 GeV it covers 3 bins. Above 30 GeV the slope decreases logarithmically with energy and crosses zero at $275\pm32$ GeV. This confirms our observation from Fig.~\ref{fig:pfrac_500} that above about 200 GeV the positron fraction is no longer increasing with energy.  This is the first experimental evidence of the existence of a new behavior of the positron fraction at high energy.

\begin{figure}[!ht]
\begin{center}
\includegraphics[width=0.68\columnwidth]{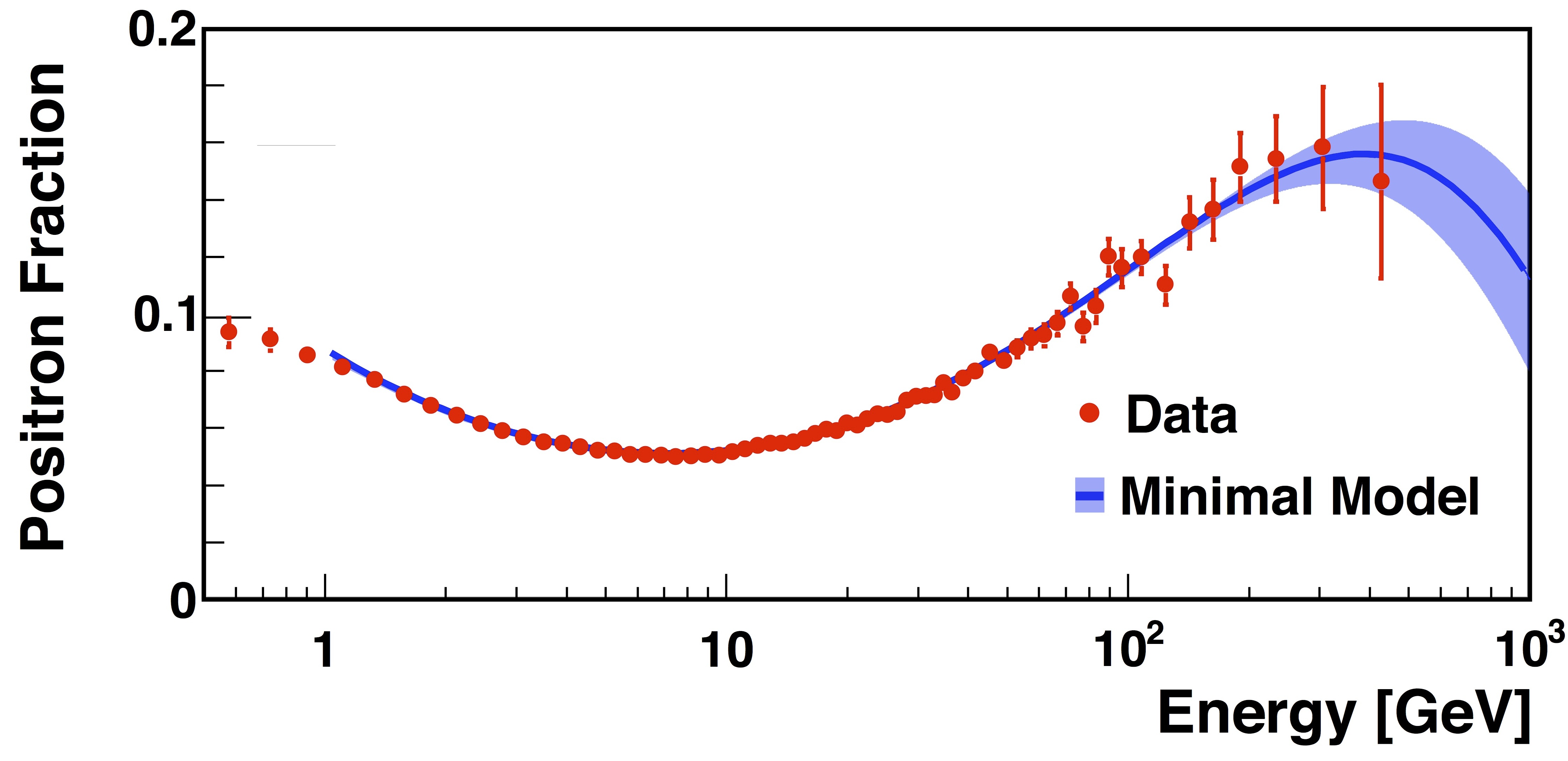} 
\caption{The positron fraction measured by AMS and the fit of a minimal model (solid curve, see text)
and the 68\% C.L. range of the fit parameters (shaded). For this fit both the data and the model
are integrated over the bin width. The error bars are the quadratic sum of the statistical and
systematic uncertainties. Horizontally, the points are placed at the center of each bin.}
\label{fig:pfrac_fit}
\end{center}
\end{figure}

We present a fit to the data of a minimal model,  where the e$^+$ and e$^-$ fluxes are parameterized as the sum of their individual diffuse power law spectrum and a common source term with an exponential cutoff parameter, $E_s$:
\begin{eqnarray}
\Phi_{\mathrm{e}^+} = C_{\mathrm{e}^+} E^{-\gamma_{\mathrm{e}^+}} + C_{s} E^{-\gamma_{s}} e^{-E/E_s} \\ 
\Phi_{\mathrm{e}^-} = C_{\mathrm{e}^-} E^{-\gamma_{\mathrm{e}^-}} + C_{s} E^{-\gamma_{s}} e^{-E/E_s} 
\end{eqnarray} 
(with E in GeV). A fit of this model to the data with their total errors (the quadratic sum of the statistical and systematic errors) in the energy range from 1 to 500 GeV yields a $\chi^2/\mathrm{d.f.} = 36.4/58$ and the cutoff parameter $1/E_s = 1.84 \pm 0.58$ TeV$^{-1}$, while the other parameters have similar values to those in~\cite{AMS_pf_250}, $C_{\mathrm{e}^+}/C_{\mathrm{e}^-} = 0.091 \pm 0.001$, $C_s/C_{\mathrm{e}^-} = 0.0061 \pm 0.0009$, $\gamma_{\mathrm{e}^-} - \gamma_{\mathrm{e}^+} = -0.56 \pm 0.03$, and $\gamma_{\mathrm{e}^-} - \gamma_s = 0.72 \pm 0.04$. The resulting fit is shown in Fig.~\ref{fig:pfrac_fit} as a solid curve together with the 68\% C.L. range of the fit parameters. No fine structures are observed in the data. The same model with no exponential cutoff parameter, i.e. with $1/E_s$ set to 0, is excluded at the 99.9\% C.L. 

In~\cite{AMS_pf_250} we reported that solar modulation has no observable effect on our measured positron fraction
and this continues to be the case. An analysis of the arrival directions of positrons and electrons was performed including the additional data. The positron to electron ratio remains consistent with isotropy; the upper limit on the amplitude of a potential dipole anisotropy
is $\delta \leq 0.030$ at the 95\% C.L. for energies above 16 GeV.

\section{Summary}

The single flux measurements, their sum and in particular the positron ratio make possible the accurate comparison with various particle physics and astrophysics models including the minimal model discussed above. The latter comparison will be presented in a separate forthcoming publication. The new improved measurement of the positron fraction shows a previously unobserved behavior. Above about 200 GeV, the positron fraction no longer increases. This, as well as the differing behavior of the spectral indices vs.~energy for electrons and positrons indicates that high energy positrons have a different origin from that of electrons.  Following the publication of our first paper~\cite{AMS_pf_250}, there have been many interesting interpretations~\cite{DM_fits} with two popular classes. In the first, the excess of e$^+ $ comes from pulsars. In this case, after flattening out with energy the positron fraction is expected to slowly decrease and a dipole anisotropy should be observed. In the second class of interpretations, the shape of the positron fraction is due to dark matter collisions. In this case, after flattening out, the fraction will decrease rapidly with energy due to the finite and specific mass of the dark matter particle and no dipole anisotropy will be observed. Over its lifetime, AMS will reach a dipole anisotropy
sensitivity of $\delta < 0.01$ at the 95\% C.L.

The underlying mechanism of this behavior can only be ascertained by continuing to collect data up to the TeV region (currently, the largest uncertainties above 200 GeV are the statistical errors) and by measuring the anti-proton to proton ratio to high energies. These are among the main goals of AMS.

\bigskip
\section{Acknowledgments}
The University of Geneva AMS group gratefully acknowledges financial support by the Swiss National Science Foundation as well as Federal and Cantonal authorities.

%
%

%
%
%
%
 
\end{document}